\documentclass[12pt]{article}
\pdfoutput=1

\usepackage[utf8]{inputenc}
\usepackage{graphics,amssymb,float}
\usepackage{graphicx}
\usepackage{color}
\usepackage{amsmath}
\usepackage{xspace}
\usepackage{placeins}
\usepackage{dcolumn}
\usepackage{bm}

\usepackage{cite}
\usepackage{hyperref}
\usepackage{indentfirst}  
\usepackage{rotating} 
\usepackage{longtable,multirow}

\def\lsim{\mathrel{\raise.3ex\hbox{$<$\kern-.75em\lower1ex\hbox{$\sim$}}}}
\def\gsim{\mathrel{\raise.3ex\hbox{$>$\kern-.75em\lower1ex\hbox{$\sim$}}}}
\def\ifmath#1{\relax\ifmmode #1\else $#1$\fi}

\def\smodels{{\sc SModelS}}

\newcommand{\met}{\ensuremath{E_{T}^{\text{miss}}}}
\begin{document}

\begin{center}

\vspace*{1cm}

{\Large\bf SModelS extension with the\\[2mm]  CMS supersymmetry search results from Run~2} 

\vspace*{1cm}

\renewcommand{\thefootnote}{\fnsymbol{footnote}}

{\large 
Juhi~Dutta$^{1}$\footnote[1]{Email: juhidutta@hri.res.in},
Sabine~Kraml$^{2}$\footnote[2]{Email: sabine.kraml@lpsc.in2p3.fr},
Andre~Lessa$^{3}$\footnote[3]{Email: andre.lessa@ufabc.edu.br}, 
Wolfgang~Waltenberger$^{4}$\footnote[4]{Email: wolfgang.waltenberger@oeaw.ac.at}
} 

\renewcommand{\thefootnote}{\arabic{footnote}}

\vspace*{1cm} 

{\normalsize \it 
$^1\,$Regional Centre for Accelerator-based Particle Physics,
Harish-Chandra Research Institute, HBNI, Chhatnag Road, Jhusi, Allahabad-211019, India\\[2mm]
$^2\,$Laboratoire de Physique Subatomique et de Cosmologie, Universit\'e Grenoble-Alpes,
CNRS/IN2P3, 53 Avenue des Martyrs, F-38026 Grenoble, France\\[2mm]
$^3\,$Centro de Ci\^encias Naturais e Humanas, Universidade Federal do ABC,\\ Santo Andr\'e, 09210-580 SP, Brazil\\[2mm]
$^4\,$Institut f\"ur Hochenergiephysik,  \"Osterreichische Akademie der Wissenschaften,\\ Nikolsdorfer Gasse 18, 1050 Wien, Austria\\[2mm]
}

\vspace{6mm}

\begin{abstract}
We present the update of the SModelS database with the simplified model results from CMS searches 
for supersymmetry at Run~2 with 36~fb$^{-1}$ of data. The constraining power of these 
new results is compared to that of the 8~TeV results within the context of a full model,
the pMSSM. The new database, v1.1.2, is publicly available and can readily be employed for physics studies with SModelS.
\end{abstract}

\end{center}


\section{Introduction}\label{sec:intro}

Simplified models~\cite{hep-ph/0703088, 0810.3921, 1105.2838, Okawa:2011xg, 1301.2175} have become one of the standard methods 
for ATLAS and CMS to optimise analyses for specific signatures, compare the reach, and communicate the results of their searches for new particles. 
When simplified model results are provided in terms of cross section upper limits or efficiency maps, they can readily be re-used to constrain arbitrary 
beyond-the-standard-model (BSM) theories in which the same final state occurs, as long as  differences in the event kinematics 
(e.g., from different production mechanisms or from the spin of the BSM particle)
do not significantly affect the signal acceptance of the experimental analysis. 
This is precisely the idea behind SModelS~\cite{Kraml:2013mwa,Kraml:2014sna}.

\begin{figure}[t!]\centering
\includegraphics[width=0.84\textwidth]{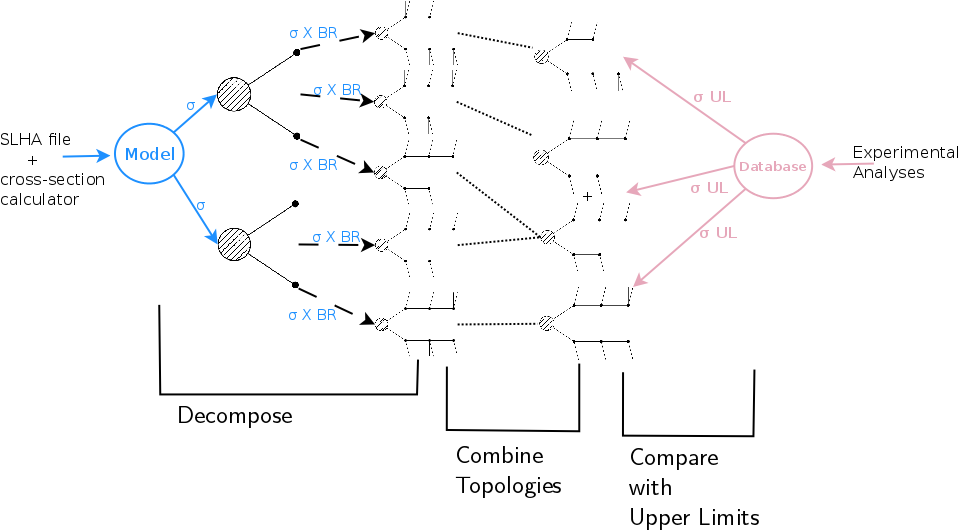}
\caption{\label{fig:smodels_scheme}
Schematic view of the working principle of SModelS.}
\end{figure}

SModelS is a public tool which allows to 
exploit the plethora of constraints on simplified model spectra (SMS) from ATLAS and CMS searches for supersymmetry (SUSY) 
in an automatised way. 
The principle of SModelS, in the current version~1.1, is to decompose BSM collider 
signatures featuring a $\mathbb{Z}_2$ symmetry into simplified model topologies, using a generic
procedure where each SMS is defined by the vertex structure and the SM final
state particles; BSM particles are described only by their masses, production
cross sections and branching ratios. The working principle is illustrated in Fig.~\ref{fig:smodels_scheme}. 
The SModelS code and database are publicly available on GitHub at \url{https://github.com/SModelS/}
or on the SModelS wiki page, \url{http://smodels.hephy.at/}.

The previous database version\cite{Ambrogi:2017neo} (v1.1.1) was comprised of 186 results (125 upper limits and 61 efficiency maps) 
from 21 ATLAS and 23 CMS SUSY searches, covering a total of 37 simplified models. 
From these 44 searches, the vast majority were based on Run~1 data. 
Only 11 (4 from ATLAS and 7 from CMS) were based on early 13~TeV Run~2 data  with 2--13~fb$^{-1}$ of integrated luminosity; 
most of these were preliminary results from ATLAS conference notes or CMS public analysis summaries. 

In this note we now present the implementation of the Run~2 SUSY search results from CMS with 36~fb$^{-1}$, presented 
at the Moriond and the summer (LHCP and EPS) conferences of 2017. 
This extends the SModelS database by 84 new cross section upper limit (UL) maps from 19 different analyses.  
We give an overview which results have been included, 
show their validation in SModelS, 
and demonstrate their constraining power for the phenomenological Minimal Supersymmetric Standard Model (pMSSM) 
as compared to the 8~TeV data. 

\clearpage

\section{CMS 13 TeV results for 36/fb included in this release}\label{sec:database}

The v1.1.2 database presented here includes results from 19 CMS SUSY analyses
from Run~2 with 36~fb$^{-1}$ of data, comprising in total 84 new SMS results for the full 2016 dataset.%
\footnote{Analogous results from ATLAS are available on HEPData and will  be added as soon as possible.} 
A detailed list is given in Table~\ref{tab:CMSAnalyses}.\\

\begin{table}[h!]\centering
\begin{tabular}{l | lcll}
&{\bf Analysis}              & {\bf Ref.} & {\bf ~~~~ID} & {\bf SMS results (txnames)} \\ \hline
\parbox[t]{3mm}{\multirow{12}{*}{\rotatebox[origin=c]{90}{Gluino, Squark}}} 
& jet multiplicity + $H_T^{\rm miss}$  & \cite{Sirunyan:2017cwe} & SUS-16-033 & T1, T1bbbb, T1tttt, \\
      & & & & T2, T2bb, T2tt\\
& jets + \met, $M_{T2}$ & \cite{Sirunyan:2017kqq} & SUS-16-036 & T1, T1bbbb, T1tttt, \\
      & & & & T2, T2bb, T2tt, T2cc, \\ 
      & & & & T6bbWW$^\dagger$\\
& 1 lept. + jets + \met, $M_J$ & \cite{Sirunyan:2017fsj} & SUS-16-037 & T1tttt, T5tttt$^\dagger$\\
& 1 lept. + jets + \met, $\Delta\Phi$ & \cite{Sirunyan:2017mrs} & SUS-16-042 & T1tttt, T5WW$^\dagger$\\
& 2 OS lept. + jets + \met  & \cite{Sirunyan:2017qaj} & SUS-16-034 & T5ZZ$^\dagger$, TChiWZ \\
& 2 SS lept. + jets + \met  & \cite{Sirunyan:2017uyt} & SUS-16-035 & T1tttt, T5WW$^\dagger$, T5ttbbWW$^\dagger$, \\
      & & & & T5tttt$^\dagger$, T5tctc$^\dagger$, T6ttWW$^\dagger$ \\
& multi-lept. + jets + \met  & \cite{Sirunyan:2017hvp} & SUS-16-041 & T1tttt, T6HHtt$^\dagger$, T6ZZtt$^\dagger$,\\
      & & & &  T6ttWW$^\dagger$ \\
& 0 lept. + top tag  & \cite{Sirunyan:2017pjw} & SUS-16-050 & T1tttt, T2tt, T5tttt$^\dagger$, T5tctc$^\dagger$ \\
\hline
\parbox[t]{3mm}{\multirow{4}{*}{\rotatebox[origin=c]{90}{Third gen.}}} 
&0 lepton stop  & \cite{Sirunyan:2017wif} & SUS-16-049 & T2tt, T2ttC, T2cc, T6bbWW$^\dagger$ \\
&1 lepton stop  & \cite{Sirunyan:2017xse} & SUS-16-051 & T2tt, T6bbWW$^\dagger$ \\
&2 lepton stop  & \cite{Sirunyan:2017leh} & SUS-17-001 & T2tt, T6bbWW$^\dagger$ \\
& $b$ or $c$-jets + \met & \cite{Sirunyan:2017kiw} & SUS-16-032 & T2bb, T2cc \\
&soft lepton, compressed stop & \cite{CMS-PAS-SUS-16-052} & PAS-SUS-16-052 & T2bbWWoff, T6bbWWoff$^\dagger$ \\
\hline
\parbox[t]{3mm}{\multirow{5}{*}{\rotatebox[origin=c]{90}{electroweak}}}
&$WH\,(H\to b\bar b)$ + \met             & \cite{Sirunyan:2017zss} & SUS-16-043 & TChiWH \\
&multi-leptons + \met             & \cite{Sirunyan:2017lae} & SUS-16-039 & TChiWH, TChiWZ, \\
    & & & & TChiChipmSlepL, \\ 
    & & & & TChiChipmSlepStau, \\
    & & & &  TChiChipmStauStau \\
&EWK combination             & \cite{CMS-PAS-SUS-17-004} & PAS-SUS-17-004 & TChiWH, TChiWZ \\
\hline
\parbox[t]{3mm}{\multirow{3}{*}{\rotatebox[origin=c]{90}{photon}}}
& Razor + $H\to\gamma\gamma$ & \cite{Sirunyan:2017eie} & SUS-16-045 & TChiWH, T6bbHH$^\dagger$ \\
& photon + \met             & \cite{Sirunyan:2017nyt} & SUS-16-046 & T5gg, T6gg \\
& photon + $H_T$             & \cite{Sirunyan:2017yse} & SUS-16-047 & T5gg, T6gg \\    
\hline
\end{tabular}
\caption{CMS 13~TeV results for 36~fb$^{-1}$ included in this SModelS database update.
The last column lists the specific SMS results included, using
the shorthand ``txname'' notation (see text for details).
For brevity, only the on-shell results are listed, although the off-shell ones are always also included  
(e.g., T1tttt in the table effectively means T1tttt and T1ttttoff).  
The superscript $\dagger$ denotes SMS with three mass parameters, for which only one mass plane is available; 
we included them for completeness but note that they apply to the given 2D slice of parameter space only, 
not to general mass patterns.}
\label{tab:CMSAnalyses}
\end{table}

\clearpage

\begin{figure}[t]\centering
\includegraphics[width=0.48\textwidth]{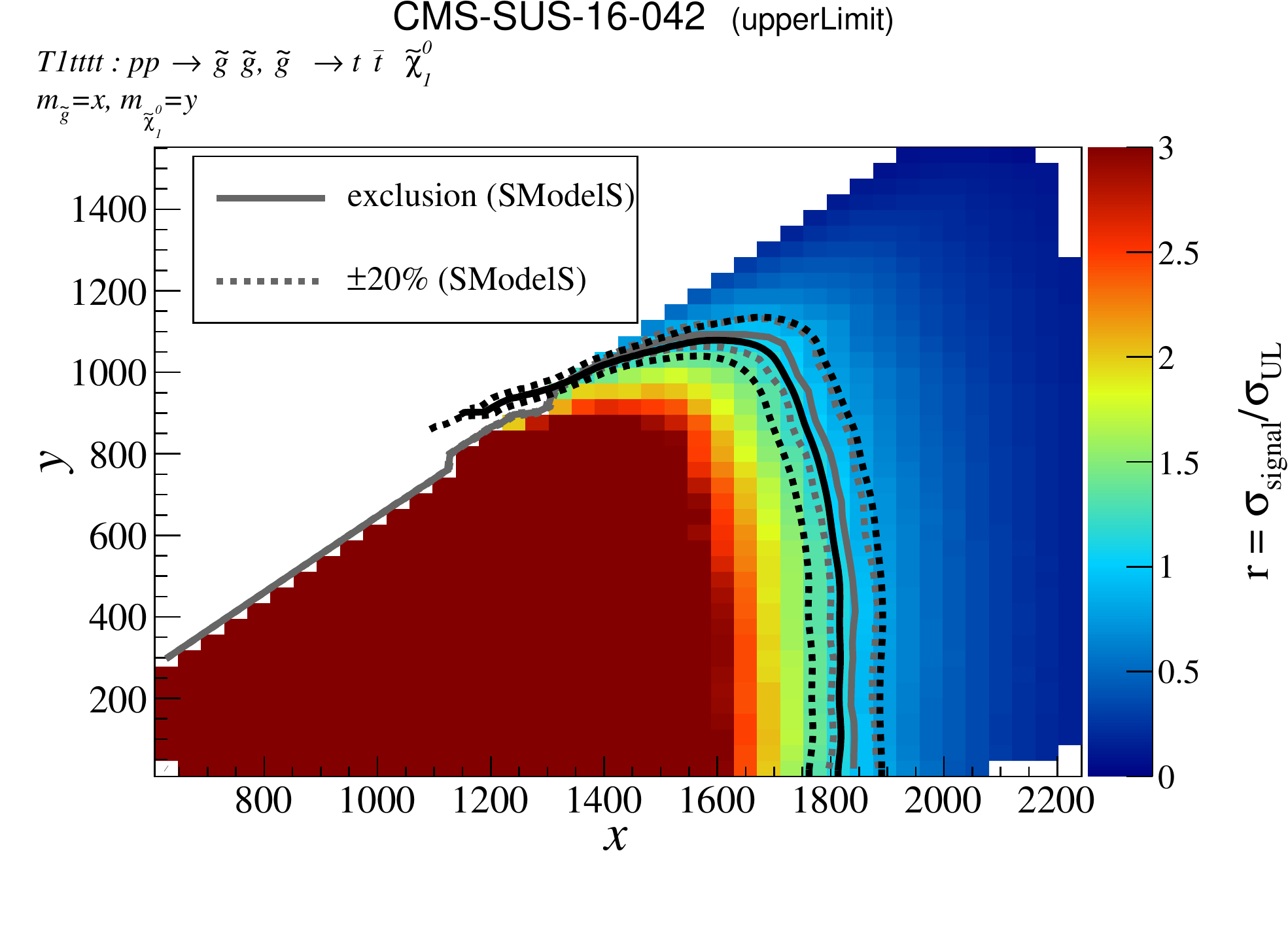}\includegraphics[width=0.48\textwidth]{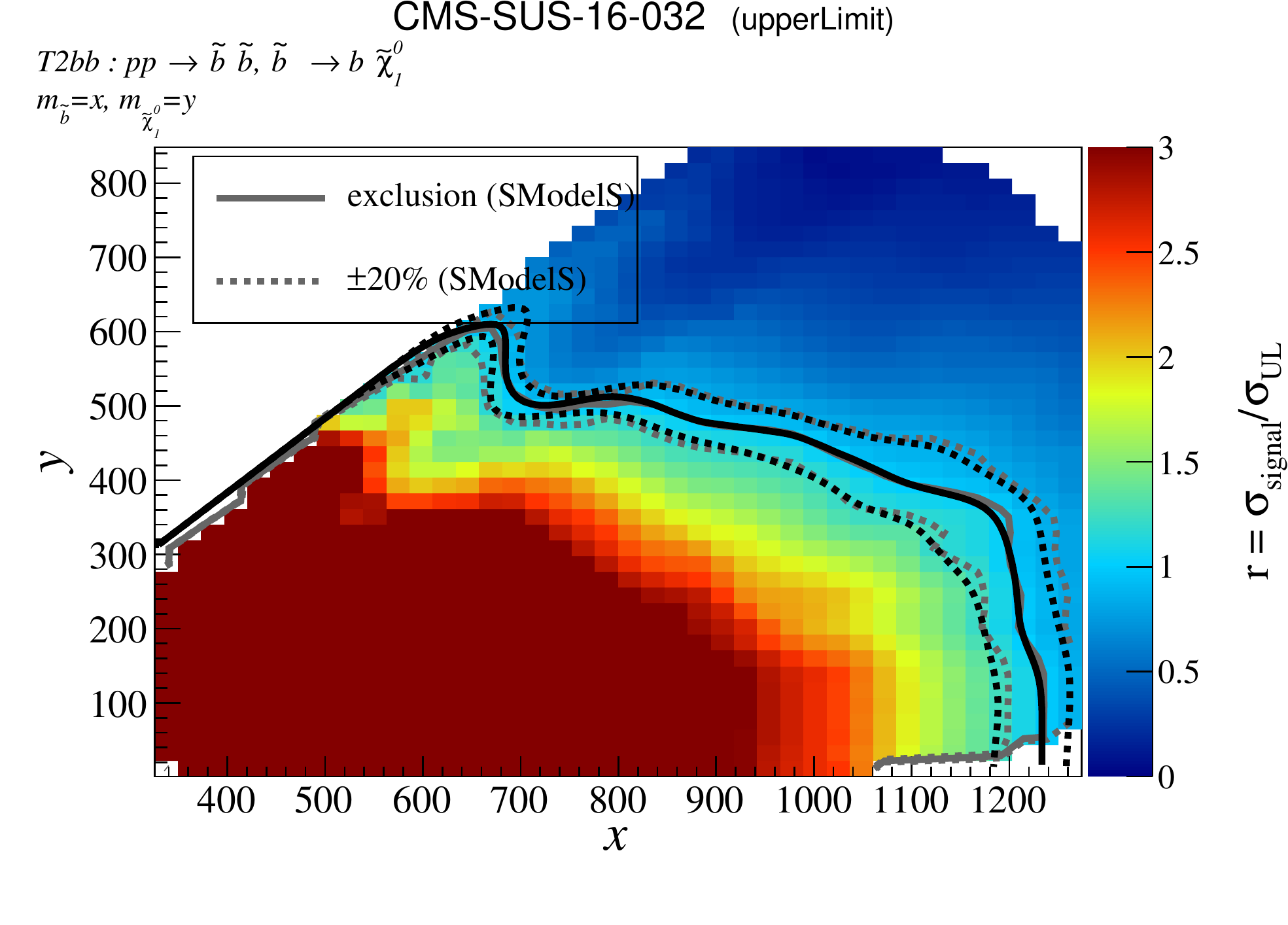}\\
\includegraphics[width=0.48\textwidth]{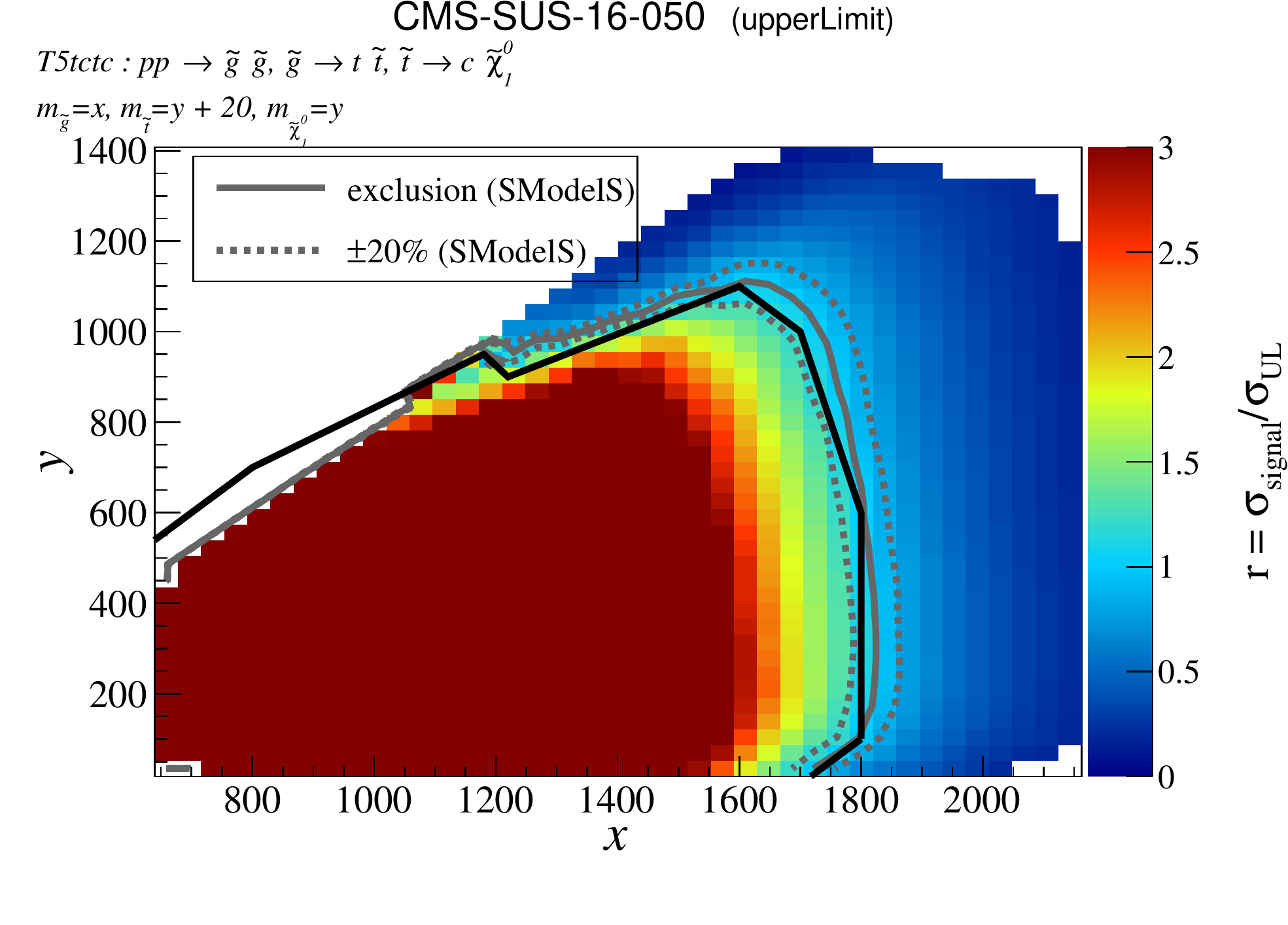}\includegraphics[width=0.48\textwidth]{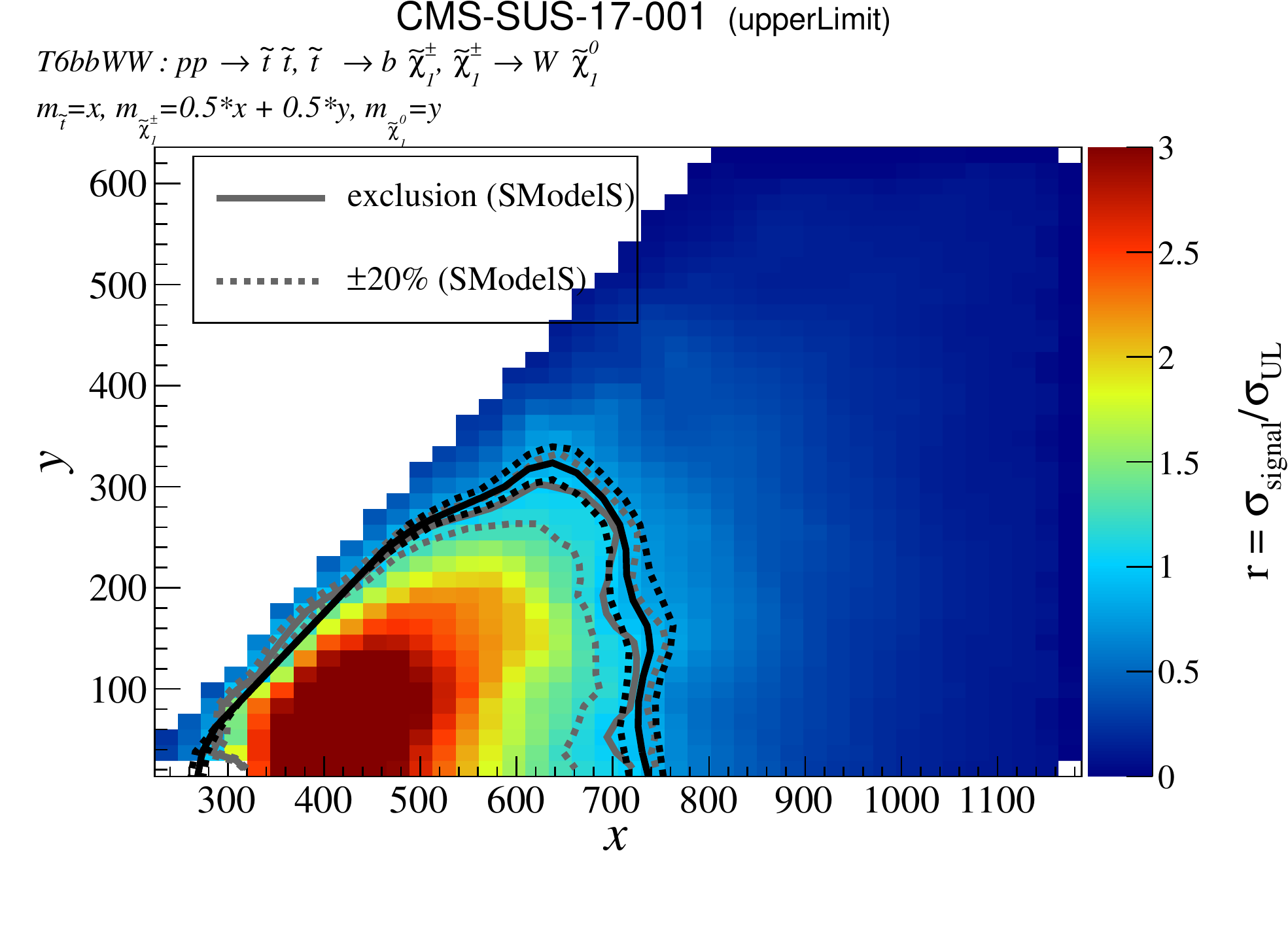}
\caption{\label{fig:validation}
Examples of validation plots. The coloured histograms show the $r$-values, defined as the ratio of the theory prediction over the observed upper limit. 
The full gray lines are the SModelS exclusion contours, $r=1$, to be compared to the official CMS exclusion lines in black ($\pm 1\sigma$ as 
dashed black lines). The effect of a 20\% variation, $r=1\pm0.2$, is indicated by the dashed grey lines.}
\end{figure}

All these new CMS results are upper limit maps: they give the 95\% confidence level (CL) upper limit values on \(\sigma
\times {\rm BR}\) for a particular SMS as a function of the relevant parameters, usually the SUSY particle masses
or slices over mass planes. They are derived from the colour maps in the simplified model limit plots 
of the experimental papers, which CMS systematically provides in numerical form, typically 
as ROOT files on the analyses' wiki pages.\footnote{Alternatively, SModelS\,v.1.1 can also use efficiency maps~\cite{Ambrogi:2017neo}. 
Efficiency maps (EMs) have the advantage that contributions from different topologies to the same signal region can be combined.} 
Each included map is thoroughly validated to make sure it reproduces the limits reported in the experimental publication. 
Figure~\ref{fig:validation} shows some examples of validation plots; the full set is available online at 
\url{http://smodels.hephy.at/wiki/Validationv112}.

Inside SModelS, individual SMS results are identified by the analysis ID and the txname (see right-most column in Table~\ref{tab:CMSAnalyses}), 
which describes in a shorthand notation the hypothesised SUSY process used to derive the UL map. 
The txnames largely follow the notation introduced in \cite{1301.2175}.  For instance, 
`T1' topologies stand for gluino-pair production followed by 3-body gluino decay into the lightest SUSY particle (LSP), 
usually the $\tilde\chi^0_1$, hence:
 \begin{equation*}
 \mbox{T1: } pp\to \tilde g\tilde g, \tilde g\to q\bar q\tilde\chi^0_1,\; \mbox{T1bbbb: } pp\to \tilde g\tilde g, \tilde g\to b\bar b\tilde\chi^0_1, 
\; \mbox{T1tttt: } pp\to \tilde g\tilde g, \tilde g\to t\bar t\tilde\chi^0_1 \ldots
 \end{equation*}
T1bbtt would mean one gluino decays into $b\bar b\tilde\chi^0_1$ and the other one into $t\bar t\tilde\chi^0_1$, 
but results for such asymmetric topologies are currently not available. `T5' also stands for gluino-pair production but with 
the decay proceeding via an intermediate onshell SUSY particle (for example T5tttt: $pp\to \tilde g\tilde g$, 
$\tilde g\to t\tilde t_1$, $\tilde t_1\to t\tilde\chi^0_1$). Along the same lines, `T2' and `T6' denote squark ($\tilde q$, $\tilde b$, $\tilde t$) 
pair production followed by, respectively, direct or cascade decay into the LSP  
(e.g., T2tt: $pp\to \tilde t\tilde t$, $\tilde t\to t\tilde\chi^0_1$; T6bbWW: $pp\to \tilde t\tilde t$, $\tilde t\to b\tilde\chi^+_1$, $\tilde\chi^+_1\to W\tilde\chi^0_1$). 
A complete list of txnames and the corresponding diagrams 
can be consulted at \url{http://smodels.hephy.at/wiki/SmsDictionary}.

We note also that, whenever relevant, the experimental results for topologies with top quarks and/or massive gauge bosons are split into 
several UL maps according to different kinematic regions where the tops, $W$s or $Z$s are on-shell or off-shell.  For example, an experimental result 
for $pp\to \tilde t\tilde t$, $\tilde t\to t\tilde\chi^0_1$ will have two UL maps in \smodels, one called T2tt covering the region where the 
$\Delta m = m_{\tilde t}-m_{\tilde\chi^0_1} \ge m_t-2\Gamma_t$, $\Gamma_t$ being the top total width, 
and one called T2ttoff covering $m_W<\Delta m<m_t-2\Gamma_t$. 
The reason is that for T2tt the final state to be constrained is $2t+\met$ while for T2ttoff it is $2b2W+\met$. 
(Below $\Delta m=m_W$, one enters a different regime of stop 4-body or loop decays.)
The $2b2W+\met$ final state also arises from stop decays via a chargino, $\tilde t\to b\tilde\chi^+_1\to bW^+\tilde\chi^0_1$, 
but this has a different topology (vertex structure) and corresponds to a distinct simplified
model (T6bbWW).
For conciseness, the ``off'' maps are not listed in Table~\ref{tab:CMSAnalyses}, with the exception of PAS-SUS-16-052, which has only 
UL maps for compressed spectra where $W$s are always off-shell. 
 
In total, the 84 new results in the v1.1.2 database cover 25 distinct topologies (35 when counting on- and off-shell versions separately). 
As can also be seen in Table~\ref{tab:CMSAnalyses}, several analyses have SMS interpretations for the same topologies (txnames). 
For instance, upper limits for $pp\to \tilde g\tilde g$, $\tilde g\to t\bar t\tilde\chi^0_1$ (T1tttt) are provided in seven of the eight searches 
for gluinos.  Likewise, the different stop searches in the 0, 1, and 2 leptons channels all give upper limits for 
$pp\to \tilde t_1\tilde t_1$, $\tilde t_1\to t\tilde\chi^0_1$ (T2tt) and $pp\to \tilde t_1\tilde t_1$, $\tilde t_1\to b\tilde\chi^+_1\to bW\tilde\chi^0_1$ (T6bbWW). 
In principle it would be possible to compile, for each topology, the limits from different analyses into one single map, 
using only the strongest constraint in each mass bin. 
Instead, we have chosen to include all the individual results 
which are provided by the experimental collaboration. 
This makes the database larger and the evaluation slightly slower, but has the advantage of more flexibility. For instance it allows 
to compare the constraining power of different analyses for the same signal. 
When speed is a limiting factor, knowledgeable users can build a slimmed-down pickle file, 
applying only the subset of analyses which give the strongest constraints; see the SModelS\,v1.1 manual~\cite{Ambrogi:2017neo} 
for more details.

There is a further reason for including all individual SMS results: 
when using SModelS to constrain non-SUSY scenarios,  it is possible that, depending on the selection cuts in the analyses, 
some SMS results do not apply.  Such results should then be disregarded. 
Generally, the validity of the SMS assumptions depends on the concrete model under consideration, as well 
as details of the experimental search. It is the responsibility of the user to verify this case by case when 
testing new theories.

\clearpage

\section{Impact on the pMSSM}\label{sec:pMSSM}

To assess the impact of these new 13~TeV results in a general manner, we make use of the extensive scan of the 
pMSSM~\cite{Djouadi:1998di} with 19 free parameters from the ATLAS pMSSM study~\cite{Aad:2015baa} (see also \cite{Berger:2008cq,CahillRowley:2012cb,CahillRowley:2012kx,Cahill-Rowley:2014twa}). 
The ATLAS collaboration made the whole scan, in total more than 310k parameter points with SUSY masses up to 4 TeV, 
publicly available on HepData~\cite{ATLASpMSSMhepdata}. 
These points were classified into three sets according to the nature of the LSP: bino-like (103,410 points), wino-like (80,233 points) and higgsino-like (126,684 points).  They all have  $m_h=[124,\,128]$~GeV and satisfy constraints from SUSY searches at LEP and the Tevatron, flavor and electroweak precision measurements, cold dark matter relic density and direct dark matter searches. 
We remove from this dataset the points which contain long lived charged sparticles ($c\tau>1$~mm), 
which cannot be treated in the official SModelS version. 
This has only a small effect on the bino-like and higgsino-like LSP sets (99,492 and 123,498 points remaining, respectively) but 
removes most of the wino-like LSP points (only 8,772 points remaining). 

For this dataset, we analyse how the SModelS exclusion improves with the new 13~TeV results as compared  
to the 8~TeV results. 
As a first overview, we list in Table~\ref{tab:smodpmssmsum} the total number of points studied, 
the number of points that can be excluded by SModelS when using only the 8~TeV results in the database, 
and the number of points that can be excluded when using the full 8~TeV + 13~TeV database. As one can see, 
the gain is quite substantial, between a factor of $2$ for the higgsino-like LSP dataset and a factor of $2.7$ for the bino-like dataset. 

\begin{table}[h!]
\centering
\begin{tabular}{l | c | c | c}
 & Bino-like LSP & Higgsino-like LSP & Wino-like LSP\\
\hline
Total number of points & 99,492 & 123,498 & 8,772 \\
\hline
\# points excluded -- 8 TeV results only & 23,253 & 32,219 & 1,389 \\
\hline
\# points excuded -- full database & 62,159 & 65,768 & 3,212
\end{tabular}
\caption{Summary of results, listing the total number of points tested from the ATLAS pMSSM scan (without long-lived charged particles), the number of points excluded by SModelS using only the 8 TeV database and the number of points excluded when using the full database with 8 TeV and 13 TeV results.}
\label{tab:smodpmssmsum}
\end{table}

\begin{figure}[t!]\centering
\includegraphics[width=0.48\textwidth]{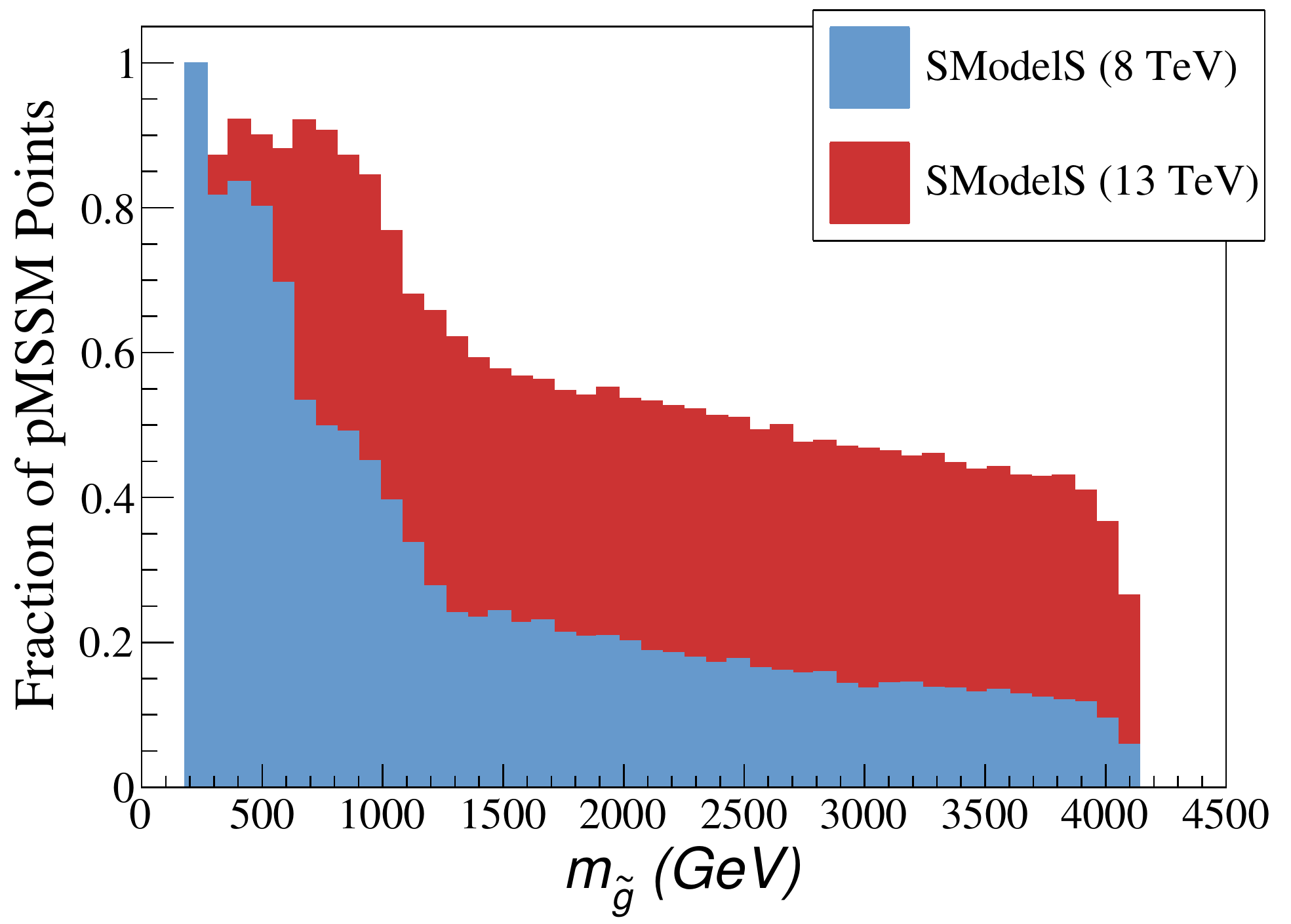}\includegraphics[width=0.48\textwidth]{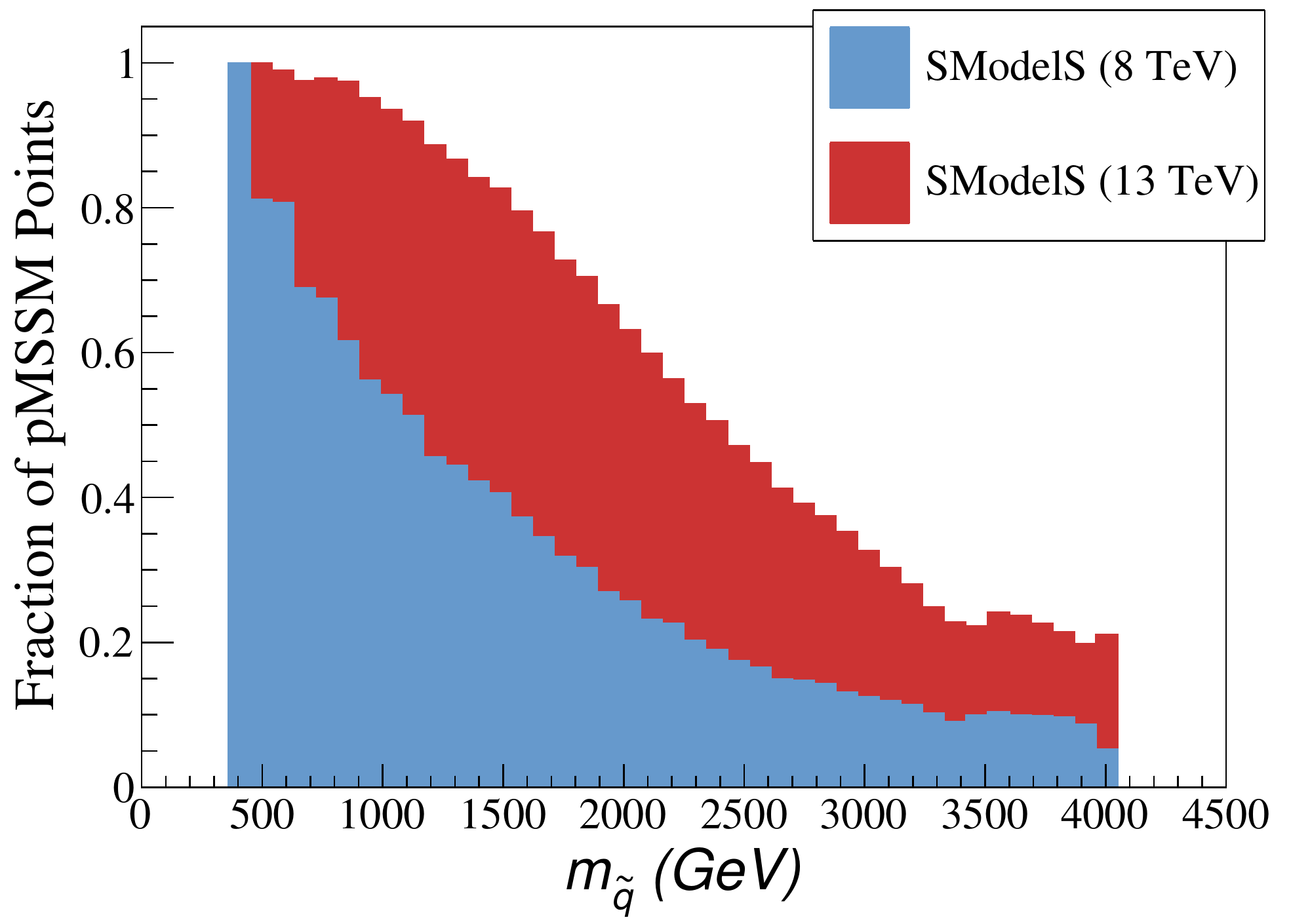}
\includegraphics[width=0.48\textwidth]{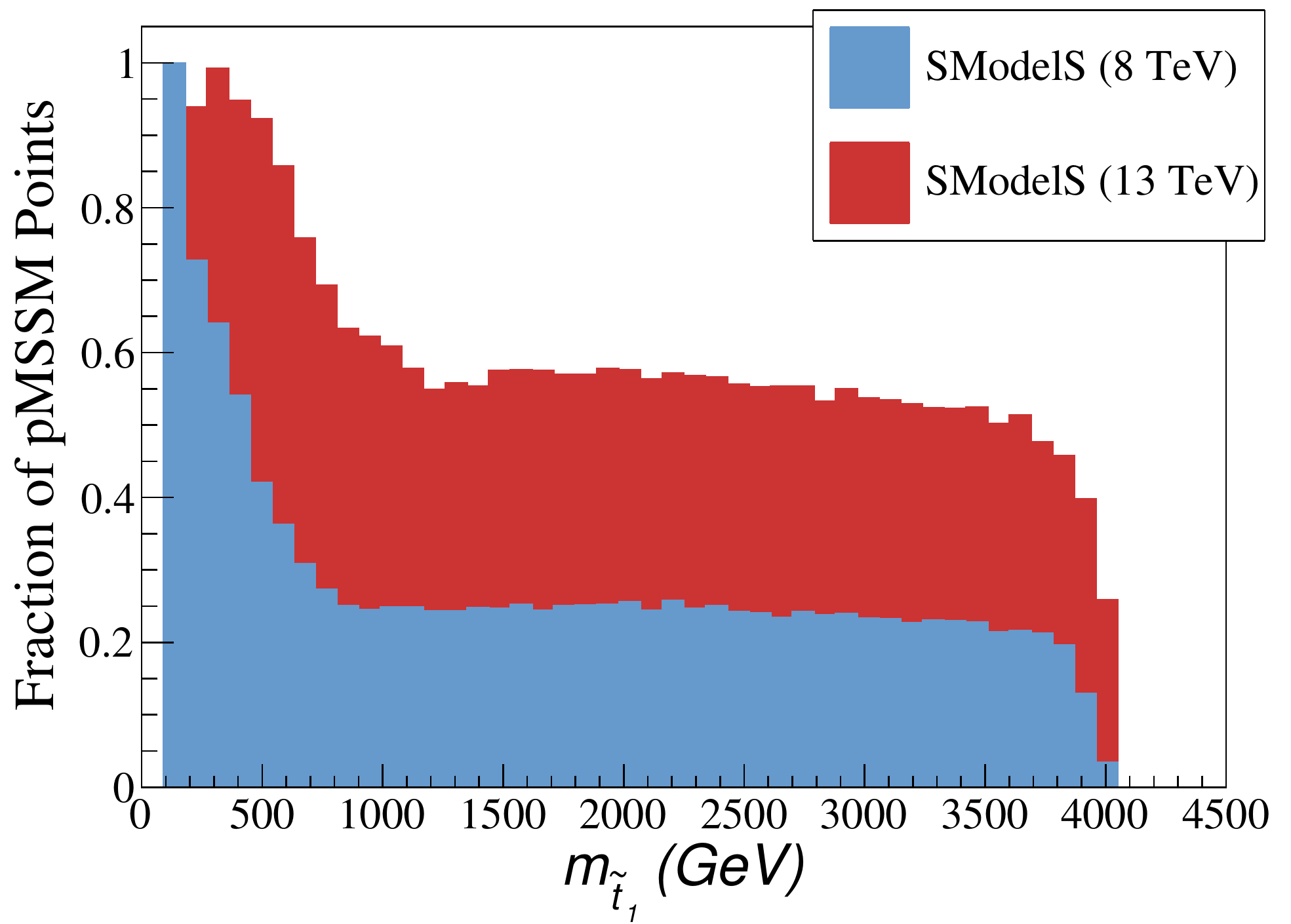}\includegraphics[width=0.48\textwidth]{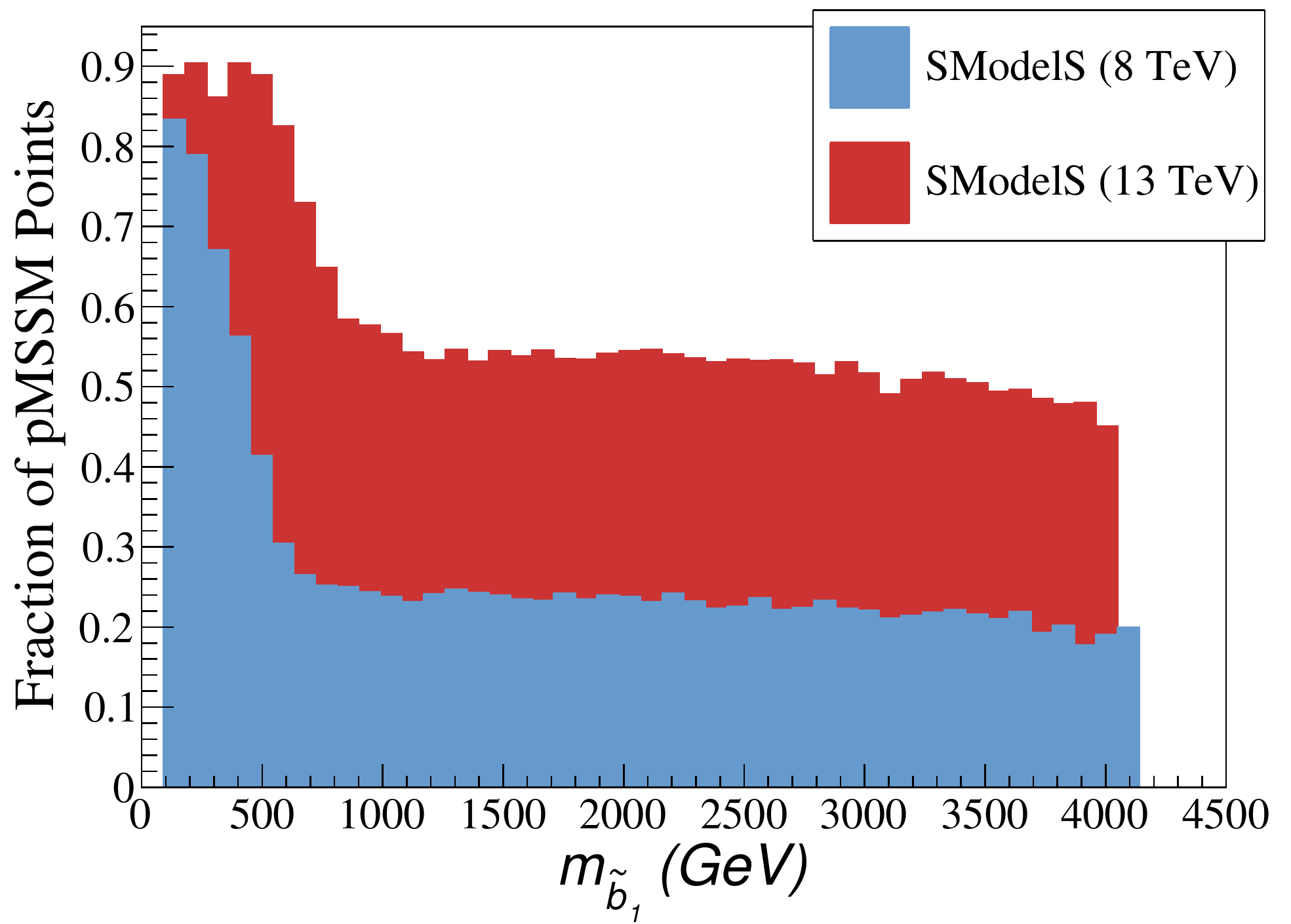}\caption{\label{fig:histexcluded}
Fraction of points excluded by SModelS for the ATLAS pMSSM scan as a function of gluino, average squark, stop and sbottom mass.
Only the points without long-lived charged particles were considered.
The blue histogram shows the fraction of excluded points using only the 8 TeV database,
while the red histogram shows the increase of excluded points once the 13 TeV database is included.}
\end{figure}

\begin{figure}[t!]\centering
\includegraphics[width=0.48\textwidth]{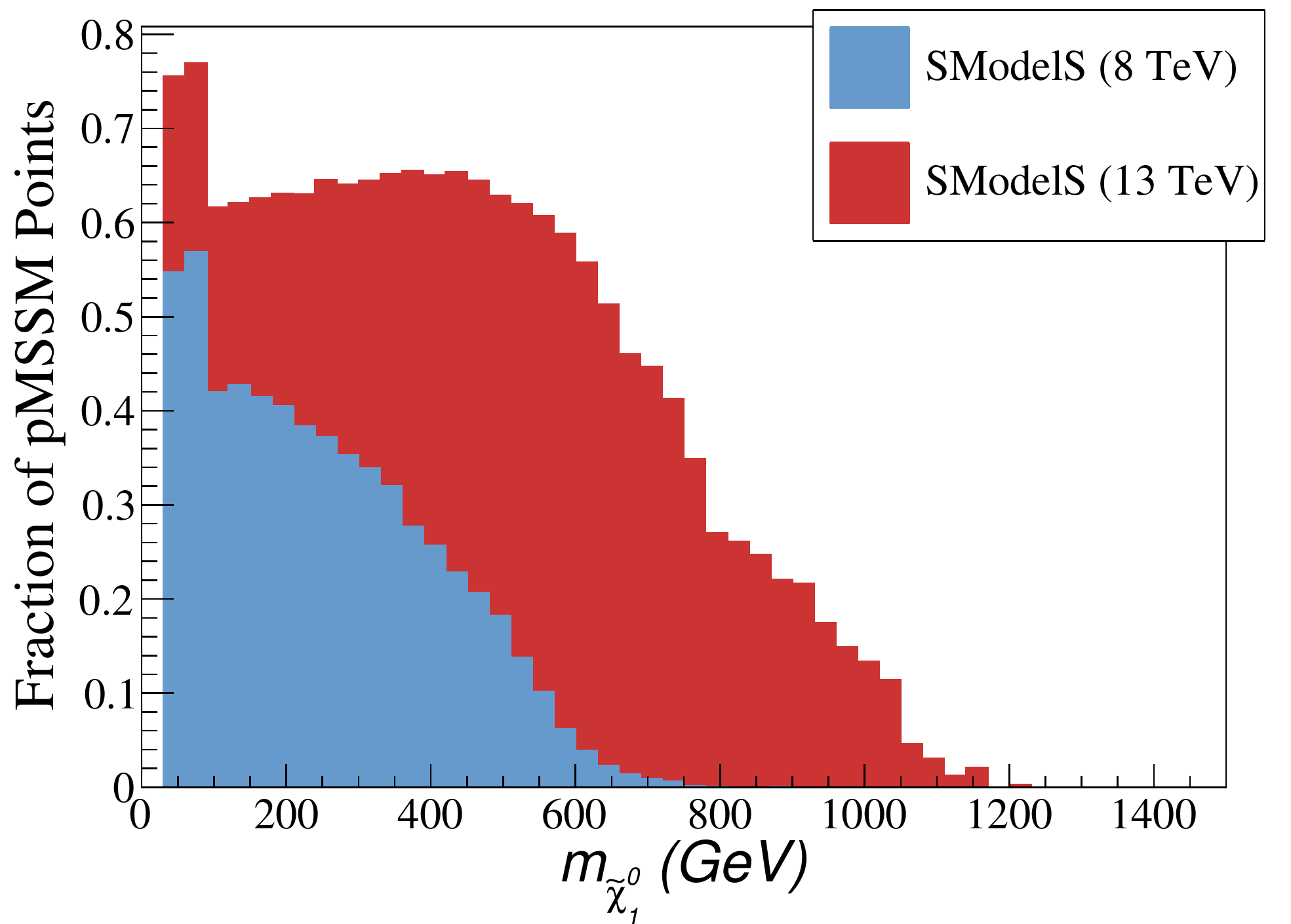}
\caption{\label{fig:histexcludedLSP}
As Fig.~\ref{fig:histexcluded} but for the neutralino LSP mass.}
\end{figure}

The impact on the gluino, average squark, stop and sbottom masses is illustrated in Fig.~\ref{fig:histexcluded}.
We see that gluinos with masses below 1 (1.5)~TeV are now much better constrained, with only about 11\%  (22\%) of points 
escaping exclusion by simplified model results in this mass range. Likewise, the SMS constraints are severely closing in 
on stops, sbottoms and light-flavor squarks, with around 70\% of points with at least
one squark below 1~TeV being excluded.
Also interesting is the impact on the LSP mass, shown in Fig.~\ref{fig:histexcludedLSP}. 
The 8~TeV results eliminate about 54\% of the pMSSM points with LSP masses below about 100~GeV, but 
show a steep drop in constraining power for heavier $\tilde\chi^0_1$. 
The new 13~TeV results, on the other hand, provide strong constraints for $\tilde\chi^0_1$ masses up to about 600~GeV, 
excluding 64\% of the pMSSM points in this range and more than 75\% of the points with $m_{\tilde\chi^0_1}\lesssim100$~GeV.

\begin{figure}[t!]\centering
\includegraphics[width=0.5\textwidth]{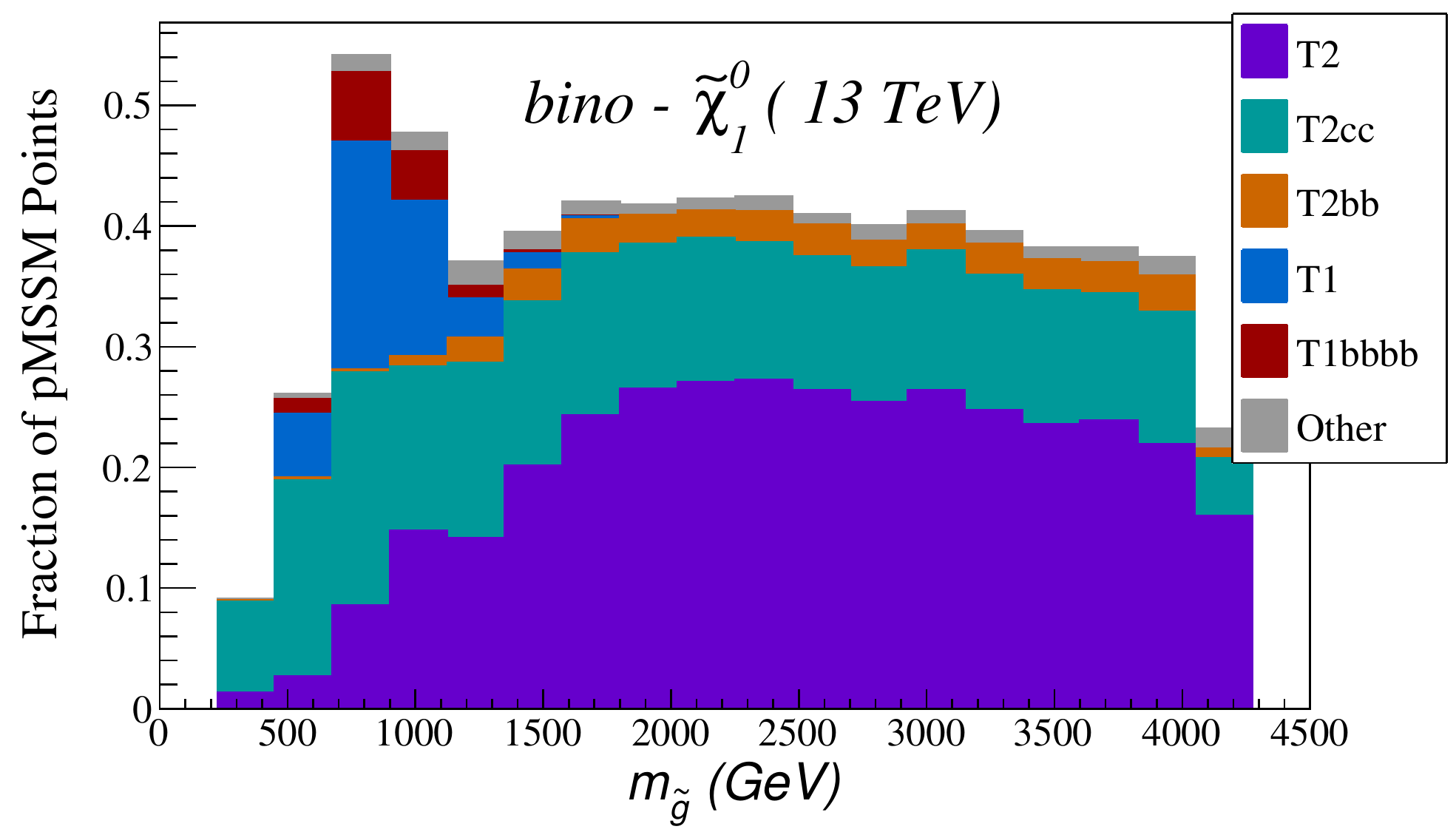}\includegraphics[width=0.5\textwidth]{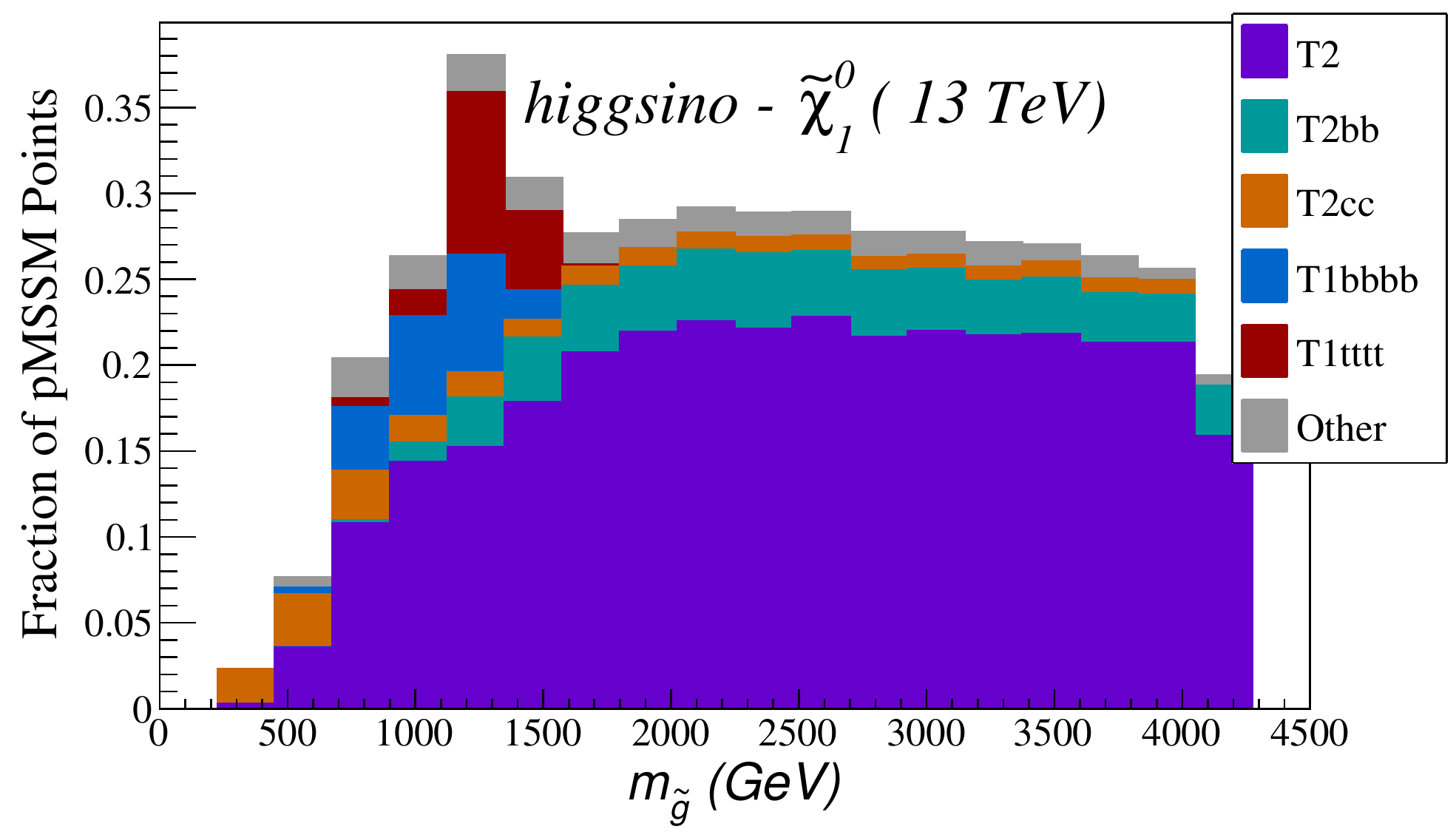}\\
\includegraphics[width=0.5\textwidth]{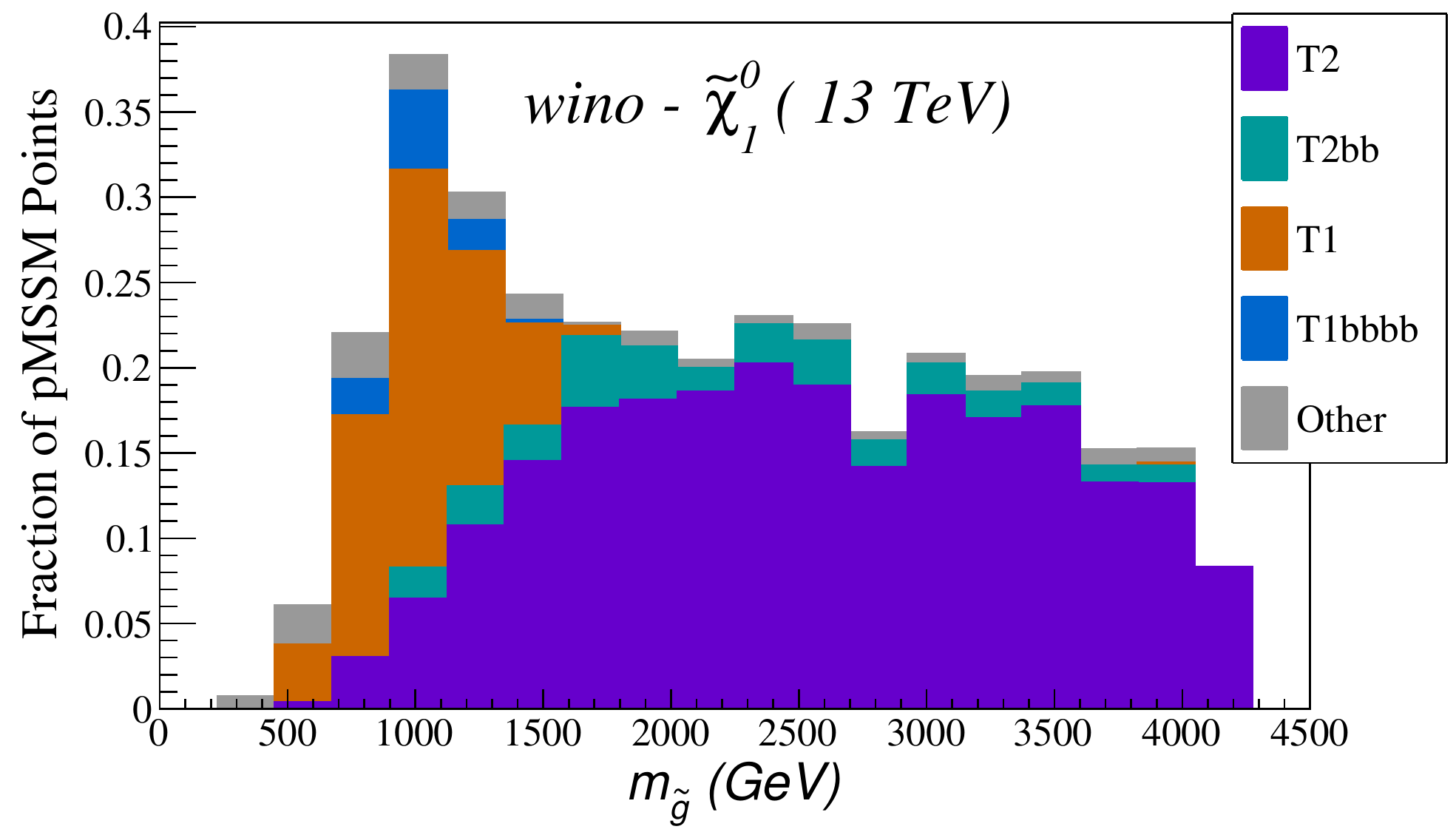}
\caption{\label{fig:txnames}
Fraction of points excluded by SModelS for the ATLAS pMSSM scan as a function of
the gluino mass for the bino-LSP, higgsino-LSP and wino-LSP scenarios. 
Only the points without long-lived charged particles were considered. 
The colored histograms show the topologies which give the highest exclusion.}
\end{figure}

To address the question which signal topologies are most relevant for the improved constraints, 
Fig.~\ref{fig:txnames} provides a break-down by txnames  
as a function of the gluino mass. 
For each point excluded at 13 TeV, but not at 8 TeV, we take the txname with the highest $r$-value ($r=\sigma_{\rm SMS}/\sigma_{\rm UL}$) 
and then show the (stacked) histograms for each txname normalized by the total number of points in each bin.

As we can see, when gluinos are within reach (i.e., for $m_{\tilde g}\lesssim1.5$~TeV), T1 and T1bbbb  
are among the most constraining topologies for the bino-like and wino-like datasets;
for higgsino-like LSP dataset gluino decays via the 3rd generation are more important 
and thus T1bbbb and T1tttt are among the most constraining topologies. 
Overall, however, and increasingly so at high gluino masses, the strongest exclusion comes from squark topologies. 
Indeed, T2 (2\,jets + $\met$) is clearly the leading topology with some contribution also from T2bb (2\,$b$-jets + $\met$). 

A comment is in order regarding the prominence of the T2cc topology in the bino-like LSP dataset. 
In principle, T2cc describes stop-pair production followed by stop decays into $c+\tilde\chi^0_1$. 
However, from the three analyses \cite{Sirunyan:2017kqq,Sirunyan:2017wif,Sirunyan:2017kiw} which provide constraints for this case, only \cite{Sirunyan:2017kiw} includes charm tagging. The other two \cite{Sirunyan:2017kqq,Sirunyan:2017wif} actually constrain 2\,jets + $\met$, not 2\,$c$-jets + $\met$, so they also apply to what is normally a T2 topology, i.e.\ $pp\to \tilde q\tilde q$, $\tilde q\to q\tilde\chi^0_1$ or  $pp\to \tilde g\tilde g$, $\tilde g\to g\tilde\chi^0_1$. 
We note here that the conventional T2 UL maps cover squark-LSP mass differences down to 25~GeV only. 
The UL maps for T2cc from \cite{Sirunyan:2017kqq,Sirunyan:2017wif}, on the other hand, are designed to cover the compressed region
and go down to mass differences of 11--12~GeV. Furthermore, for mass differences $\lesssim 80$~GeV, the T2cc results are more 
constraining than the T2 results. 
They can therefore considerably extend the exclusion of points with one light squark or gluino close in mass to the LSP. 

\section{Download and installation}\label{sec:conclusions}

The full v.1.1.2 database is now included in the SModelS\,v1.1 release available 
on GitHub at \url{https://github.com/SModelS/} or on the SModelS homepage, 
\url{http://smodels.hephy.at/}. Installation instructions 
are given in the manual, available as paper \cite{Ambrogi:2017neo} and online, and the 
{\tt INSTALLATION.rst} file in the distribution. 

For people who have already installed SModelS, {\tt smodels-database-v1.1.2.tar.gz} is also available separately from 
\begin{quote} 
\url{http://smodels.hephy.at/wiki/CodeReleases}\,.
\end{quote}
For the standard installation, it suffices to put this tarball into the main {\tt smodels} folder and explode it there.
That is, the following steps need to be performed
\begin{verbatim}
   mv smodels-database-v1.1.2.tar.gz <smodels folder>
   cd <smodels folder>
   tar -xzvf smodels-database-v1.1.2.tar.gz
   rm smodels-database-v1.1.2.tar.gz
\end{verbatim} 
The v1.1.2 database will be unpacked into the {\tt smodels-database} directory, replacing the previous version
and the pickle file will then be automatically rebuilt on the next run of SModelS.
For a clean installation, it is recommended to first remove the previous database version.
If the tarball is unpacked to another location,
one has to correctly set the SModelS database path when running SModelS.
If using {\tt runSModelS.py}, this is done in the {\tt parameters.ini} file.

Alternatively, the database can also be obtained from the
\begin{quote} 
\url{https://github.com/SModelS/smodels-database-release}
\end{quote}
repository.

\section{Conclusions and outlook}\label{sec:conclusions}

We presented the update of the SModelS database with the simplified model cross section upper limits from 19 
CMS SUSY analyses from Run~2  with 36~fb$^{-1}$ of data. These results significantly improve previously available 
constraints. Using the pMSSM as a showcase for a realistic model, we demonstrated how the limits on various SUSY masses 
are pushed to higher values by the 13~TeV results as compared to 8~TeV results. The improved constraints affect not only 
the masses of colored sparticles---particularly noticeable are the much stronger constraints on LSP masses up to about 600 GeV. 
All in all, the number of points from the ATLAS pMSSM scan~\cite{Aad:2015baa} which can be excluded by SModelS 
increases by a factor 2.3 as compared to the 8~TeV results. 

The v1.1.2 database is publicly available and can readily be used in SModelS to constrain arbitrary BSM models 
which have a $\mathbb{Z}_2$ symmetry, provided the SMS assumptions \cite{Kraml:2013mwa,Ambrogi:2017neo} apply.
The simplified model results from ATLAS searches for 36~fb$^{-1}$ at 13~TeV available on HEPData will be included as soon as possible.

\section*{Acknowledgements} 

We thank the CMS SUSY group for providing a vast amount of SMS cross section upper limits in digital format. 
Moreover, we owe special thanks to Federico Ambrogi for his contribution in the early stage of this work. 
 
J.D.\ is partially supported by funding available
from the Department of Atomic Energy, Government of India, for the Regional
Centre for Accelerator-based Particle Physics (RECAPP), Harish-Chandra Research
Institute.  She thanks moreover the LPSC Grenoble for hospitality, and 
the INFRE-HEPNET (IndoFrench Network on High Energy Physics) of CEFIPRA/IFCPAR (Indo-French Centre for the Promotion of Advanced Research), 
as well as the ``Investissements d'avenir, Labex ENIGMASS'' for financial support for a research visit in May 2017, 
during which this work was started. 

S.K.\ acknowledges support from the IN2P3 project ``Th\'eorie -- LHCiTools'' and the CNRS-FAPESP collaboration PRC275431. 
A.L.\  acknowledges support by the S\~ao Paulo Research Foundation (FAPESP), projects 2015/20570-1 and 2016/50338-6.  

\appendix
\section{Note on results not included in the v1.1.2 database}

A couple of results from the CMS publications listed in Table~\ref{tab:CMSAnalyses} have not been implemented 
in the v1.1.2 database, because they cannot be re-used well in SModelS. This is notably the case for SMS results 
with mixed decay modes, where different intermediate $\mathbb{Z}_2$-odd particles and/or different final states are summed over. 
They pose constraints on a very specific {\it sum of topologies}, which is not applicable to the general case.   
Examples are: 
\begin{itemize}
\item Fig.~12d in SUS-16-033 and Fig.~5b in SUS-16-041: these are constraints on gluino-pair production 
followed by $\tilde g\to q\bar q \tilde\chi^\pm_1\to q\bar q W\tilde\chi^0_1$ and  
         $\tilde g\to q\bar q \tilde\chi^0_2\to q\bar q Z\tilde\chi^0_1$ decays with 50\% branching ratio each. 
CMS treats this as a T5VV (V=$W,Z$) topology. For SModelS, however, this results represents an UL map 
for the weighted sum of three topologies, 25\% T5WW + 25\% T5ZZ + 50\% T5WZ. 
\item Fig.~8b in SUS-16-036, Fig.~9 in SUS-16-049 and Fig.~7 in SUS-16-051: these are limits on stop-pair production 
for ${\rm BR}(\tilde t\to b\tilde\chi^+_1)={\rm BR}(\tilde t\to t\tilde\chi^0_1)=0.5$. As in the bullet item above, the UL applies to the weighted 
sum of three topologies, 25\% T2tt + 25\% T2bb + 50\% T2tb;
\item The T5Wg results in SUS-16-046 and SUS-16-047: these are actually a sum over T5WW, T5gg and T5Wg for a given branching ratio;
\end{itemize}
As discussed in \cite{Ambrogi:2017lov}, results for asymmetric topologies, arising from two different decays happening on the two 
branches of the topology diagram, would be very useful to improve the constraining power of SMS results. 
In principle one could try to interpolate between the UL maps for the symmetric topologies with 100\% BR and the ones 
for 50\% BRs including the mixed topologies. This would  add a level of complication in the matching with 
the decomposition procedure, which is the most time-consuming part of the calculation.  
Moreover, the validation of such a procedure would  require full recasting, as there are no official results 
for intermediate BRs to compare to. 
Much better and simpler would be if efficiency maps for the individual symmetric and asymmetric topologies 
were available. This would allow to work out the limits for arbitrary branching ratios in a fast, reliable and 
robust way. 

Another class of results which are not included are long cascade decays (with more than one intermediate particle)
where intermediate masses are fixed and/or branching ratios summed over. An example is Fig.~10 of SUS-16-034. 
Here, pair-produced sbottoms decay via $\tilde b_1\to b\tilde\chi^0_2$ followed by $\tilde\chi^0_2\to l^\pm\tilde l^\mp 
\to l^+l^-\tilde\chi^0_1$ or $\tilde\chi^0_2\to Z^{(*)}\tilde\chi^0_1$. From the SModelS point of view this is not 
a simplified model topology. 

Finally, the results of a number of newer CMS publications or public analysis summaries are not included, 
because the ROOT files for the SMS limits are not yet available. This concerns the analyses 
SUS-16-048 \cite{Sirunyan:2018iwl} (2 soft leptons), SUS-17-003 \cite{CMS-PAS-SUS-17-003} (hadronic staus), 
and the searches in leptonic final states presented at the SUSY 2017 conference SUS-17-002 \cite{CMS-PAS-SUS-17-002} 
and SUS-17-009 \cite{CMS-PAS-SUS-17-009}. They will be added later when the relevant ROOT files are available.

\bibliography{references}

\end{document}